\documentclass[10pt, conference]{IEEEtran}

\usepackage{cite}
\usepackage{amsmath,amssymb,amsfonts}
\usepackage{amsthm}
\newtheorem{theorem}{Theorem}
\usepackage{algorithmic}
\usepackage{graphicx}
\usepackage{textcomp}
\usepackage{mathtools}
\usepackage{ulem}
\usepackage{xcolor}
\usepackage{algorithm,algorithmic}
\usepackage{hyperref}
\usepackage{orcidlink}
\usepackage{svg}
\usepackage{comment}
\usepackage{quantikz}
\usepackage{soul}
\usepackage{enumerate}
\usepackage{booktabs} 
\usepackage{cuted}
\usepackage{siunitx}

\newcommand\change[1]{\textcolor{red}{#1}}

\renewcommand{\ket}[2][]{{\ifthenelse{\equal{#1}{}}{|}{\underset{#1}{|}}#2\rangle}}
\renewcommand{\bra}[2][]{{\langle#2\ifthenelse{\equal{#1}{}}{|}{\underset{#1}{|}}}}
\renewcommand{\braket}[3][]{\langle #2 \ifthenelse{\equal{#1}{}}{|}{\underset{#1}{|}} #3 \rangle}

\newcommand{\ketbra}[3][]{\ifthenelse{\equal{#1}{}}{|}{\underset{#1}{|}}#2\rangle\langle #3\underset{#1}{|}}
\newcommand{\qop}[2][]{\ifthenelse{\equal{#1}{}}{\mathsf{#2}}{\underset{#1}{\mathsf{#2}}}{}}
\newcommand{\rotgs}[2][]{\qop[#1]{R}_{#2}}
\newcommand{\rotg}[3][]{\rotgs[#1]{#2}^{#3}{}}
\newcommand{\hilb}[1][]{\mathcal H\ifthenelse{\equal{#1}{}}{}{_{#1}}}
\newcommand{\reg}[1]{#1}
\newcommand{\thres}[2][]{\mathcal{T}_{#2}\ifthenelse{\equal{#1}{}}{}{\left({#1}\right)}}
\newcommand{\lthres}[2][]{\mathcal{L}_{#2}\ifthenelse{\equal{#1}{}}{}{\left({#1}\right)}}

\begin{document}
%
\title{Implementing Credit Risk Analysis with Quantum Singular Value Transformation}

\author{
    \IEEEauthorblockN{
    Davide Veronelli\IEEEauthorrefmark{1}\orcidlink{0000-0002-3810-6287},
    Francesca Cibrario\IEEEauthorrefmark{1}\orcidlink{0009-0007-8290-4992},
    Emanuele Dri\IEEEauthorrefmark{2}\orcidlink{0000-0002-5144-1514},
    Valeria Zaffaroni\IEEEauthorrefmark{1}\orcidlink{0009-0002-8221-5904},
    Giacomo Ranieri\IEEEauthorrefmark{1}\orcidlink{0009-0005-0488-2121},\\
    Davide Corbelletto\IEEEauthorrefmark{1}\orcidlink{0009-0003-8830-2619},
    Bartolomeo Montrucchio\IEEEauthorrefmark{3}
    }
    \\
    \IEEEauthorblockA{
    \IEEEauthorrefmark{1}Intesa Sanpaolo, Torino, Italy\\
    davide.veronelli@intesasanpaolo.com,
    francesca.cibrario@intesasanpaolo.com, 
    valeria.zaffaroni@intesasanpaolo.com,\\
    giacomo.ranieri@intesasanpaolo.com,
    davide.corbelletto@intesasanpaolo.com\\
     \IEEEauthorrefmark{2}Fondazione LINKS, Torino, Italy\\
    emanuele.dri@linksfoundation.com\\
    \IEEEauthorrefmark{3}DAUIN, Politecnico di Torino, Torino, Italy\\
    bartolomeo.montrucchio@polito.it
    }
}

\maketitle

\thispagestyle{plain}
\pagestyle{plain}

\begin{abstract}
The analysis of credit risk is crucial for the efficient operation of financial institutions. 
Quantum Amplitude Estimation (QAE) offers the potential for a quadratic speed-up over classical methods used to estimate metrics such as Value at Risk (VaR) and Conditional Value at Risk (CVaR). 
However, numerous limitations remain in efficiently scaling the implementation of quantum circuits that solve these estimation problems.
One of the main challenges is the use of costly and restrictive arithmetic that must be implemented within the quantum circuit. 
In this paper, we propose using Quantum Singular Value Transformation (QSVT) to significantly reduce the cost of implementing the state preparation operator, which underlies QAE for credit risk analysis. 
We also present an end-to-end code implementation and the results of a simulation study to validate the proposed approach and demonstrate its benefits.
\end{abstract}


%
\IEEEpeerreviewmaketitle

\section{Introduction}
The financial sector encompasses numerous problems of significant complexity. 
Quantum computation could potentially offer advantages when applied to such problems \cite{Ors2019, Egger2020, Herman2023} and financial institutions are beginning to show interest \cite{Sotelo2024}.
Domains such as portfolio optimization \cite{Kerenidis2019, Mugel2022, Hegade2022,Rebentrost2024, Mattesi2024}, derivative pricing \cite{Rebentrost2018, Martin2021, Chakrabarti2021, Cibrario2024}, and risk analysis \cite{Woerner2019, Egger2021, Dri2022, Dri2023} would be among the beneficiaries in the event of successful advances in quantum computing.
Credit risk analysis is a particularly pertinent example of a complex problem that could benefit from the use of quantum computing. 
However, it also serves as an example of how much progress is still needed to reach the convergence point between hardware capabilities and algorithmic optimization, which would allow large-scale applications of quantum methods to useful applications.

In the classical computing domain, the state-of-the-art involves the use of Monte Carlo simulations to determine crucial metrics for credit risk analysis, such as Value at Risk (VaR) and Expected Shortfall (or Conditional Value at Risk) \cite{credit_risk_2008, value_at_risk_2006, montecarlo_value_at_risk_2014}.
These are metrics which, together with the expected portfolio loss, enable the estimation of the so-called Economic Capital (EC), a quantity that large financial institutions are required to calculate to ensure sufficient capital reserves for solvency under adverse conditions.

Thus, in the field of financial risk management, Monte Carlo methods are an essential tool for estimating crucial expectation values. 
Quantum Amplitude Estimation (QAE) offers the possibility of obtaining the same precision with a theoretic quadratic speed-up \cite{Montanaro2015}. 
Even if asymptotically this holds true, the actual cost for implementing the needed quantum operators can be prohibitive, especially on currently available hardware. 

To address this challenge, this work investigates alternative strategies to reduce the computational overhead required for implementing QAE in credit risk analysis on NISQ (Noisy Intermediate-Scale Quantum) devices. 
In this context, we introduce a novel end-to-end implementation of a quantum algorithm for estimating VaR. 
Our primary innovation builds on the extension proposed in \cite{Dri2022, Dri2023}, which improves the original model \cite{Egger2021} to accommodate multiple risk factors in the uncertainty framework. 
While their extension effectively circumvented the use of intricate quantum arithmetic, it led to an exponential growth in the number of gates required to encode the problem.
In this work, we propose an amplitude loading approach that does not require costly arithmetic while achieving better scaling with respect to the number of counterparties taken into consideration.
Drawing inspiration from the work in \cite{Stamatopoulos2024, mcardle2022quantumstatepreparationcoherent}, our implementation uses Quantum Singular Value Transformation (QSVT) \cite{Gilyen2019} to improve the overall scaling rate.

The rest of this paper is organized as follows. Section \ref{sec:QAE_and_A} and \ref{sec:qsvt} provide the necessary preliminaries for the proposed method detailed in Section \ref{sec:methodology}. Then, Section \ref{sec:results} presents the results of the simulation experiments conducted to validate our approach. Finally, Section \ref{sec:conclusion} concludes the paper presenting the final remarks and potential future works.

\section{Quantum Risk Analysis}
\label{sec:QAE_and_A}
In recent years, several proposals have been advanced for credit risk analysis using quantum algorithms. A common feature among these proposals is the core reliance on the quantum amplitude estimation (QAE) algorithm \cite{Woerner2019, Egger2021}. Therefore, we provide an overview of the key concepts related to QAE, as well as the fundamental state preparation operator $\mathcal{A}$, which encodes the original problem in a quantum circuit.

\subsection{Quantum Amplitude Estimation}
In \cite{Brassard2002}, the authors firstly generalize Grover's algorithm, introducing Quantum Amplitude Amplification (QAA) and then combine this routine with concepts from Shor's algorithm, namely Quantum Phase Estimation (QPE).
The result, as demonstrated in \cite{Montanaro2015, Wocjan2009}, is a powerful quantum algorithm that can offer an asymptotically quadratic speed-up over the traditional Monte Carlo methodology.
	
        
To be more specific, let us assume that we are able to map our problem to a quantum operator $\mathcal{A}$ such that
\begin{equation}
\label{eq:sp_task}
\mathcal{A}\ket{0}_{n+1}=\sqrt{1-a}\ket{\psi_0}_n\ket{0} +\sqrt{a}\ket{\psi_1}_n\ket{1} 
\end{equation}
and that $\mathcal{A}$ has an inverse. We can amplify $a$ (the normalized probability of interest related to the good states, marked by $\ket{1}$) by applying powers of a specific unitary operator to $\mathcal{A}\ket{0}_{n+1}$. In particular, the operator is as follows:
        \begin{equation}
        \mathbf{G} = \mathcal{A}S_0\mathcal{A}^{\dag}S_{\chi}
    \end{equation}
Where $S_0$ changes the sign of the amplitude whenever the state is in the zero state, while $S_{\chi}$ flips the sign of the amplitudes of the good states \cite{Brassard2002}.

Traditionally, the procedure to estimate $a$ requires the use of QPE. 
For this reason, QAE needs an additional \textit{counting} register of $m$  evaluation qubits in order to perform $2^m$ applications of $\mathbf{G}$. 
The counting register, initially put in an equal superposition state by Hadamard gates, is used to control different powers of $\mathbf{G}$. After applying the inverse Quantum Fourier Transform, their state is measured resulting in an integer which is then classically mapped to the estimator $\tilde a$.

However, in recent years, several works have proposed alternative versions of QAE that use only Grover iterations, removing the need for QPE and, consequently, the additional quantum register.
The advantage of these approaches is a reduction in terms of resources needed for practical applications, both in terms of qubits and circuit depth. 
Among the most promising methods is the one cited in \cite{Suzuki2020}, which relies on replacing the QPE subroutine with a combination of simple Grover iterations and Maximum Likelihood Estimation. 
Another notable method is the Iterative Quantum Amplitude Estimation (IQAE) described in \cite{Grinko2021}, which requires iterative queries to the QPU (Quantum Processing Unit) and is the approach adopted for the experiments whose results are presented in Section \ref{sec:results}.

\subsection{State Preparation}
Since QAE is the core subroutine of the quantum risk analysis algorithm, an efficient design for the operator $\mathcal{A}$, also known as \textit{state preparation}, is crucial for its successful large-scale implementation.

In the context of VaR estimation, QAE is used to evaluate $F$, the cumulative distribution function (CDF) of losses, given a target loss value. 
This enables estimating VaR by searching for the loss value providing the desired confidence level $\alpha_{var}\in(0,1)$:
\begin{equation}
\text{VaR}_{\alpha_{var}}(L) = \inf \{ x \in\mathbb R \mid F_x(L) \geq 1-\alpha_{var} \}
\end{equation}

Therefore, the state preparation should encode in $a$ (refer to Equation \ref{eq:sp_task}) the proper value of the CDF.

In the original implementation \cite{Woerner2019, Egger2021}, the state preparation operator consists of three sub-operators: $\mathcal{U}$, $\mathcal{S}$, and $\mathcal{C}$. 
The operator $\mathcal{U}$ is responsible for loading the uncertainty model into the quantum circuit. The uncertainty model is based on the Gaussian conditional independence model, which uses a single normally distributed risk factor. 
The operator $\mathcal{S}$ is an operator that performs the weighted sum to derive the total loss by aggregating that of the defaulted counterparties.
Lastly, $\mathcal{C}$ is a comparator that flips a target qubit if the total loss is less than or equal to a predefined threshold value.
More details on the implementation and associated complexity of these operators can be found in \cite{Egger2021}.

References \cite{Dri2022, Dri2023} extended the original proposals by allowing the consideration of multiple risk factors that could be correlated with each other. 
This enabled the implementation of uncertainty models that are more aligned with those currently used by major financial institutions. 
The same algorithm also avoided the use of the sum operator ($\mathcal{S}$) in favor of directly operating on amplitudes, avoiding the need for a sum register and taking a step toward the removal of the quantum arithmetic module.
However, the proposed modifications cause the number of gates required to implement $\mathcal{A}$
to scale exponentially with the number of counterparties. 
This is because each loss scenario—i.e., every possible combination of defaulted counterparties—requires a specific rotation controlled by the entire counterparties register.
Thus, the circuits quickly reach unfeasible depths for execution on NISQ hardware, especially for realistic cases.

Recently, many studies have focused on QSVT \cite{Gilyen2019}, which has demonstrated the potential to enhance the efficiency of implementing numerous algorithms, particularly when they involve costly arithmetic operations.
Since this method will be key for the algorithm proposed in this paper, the following Section is devoted to providing the reader with the essential understanding of the main idea and the potential benefits of this approach.

\section{Quantum Singular Value Transformation}
\label{sec:qsvt}

Quantum Singular Value Transformation (QSVT) arises as an extension of Quantum Signal Processing (QSP) \cite{Martin2021}, a technique inspired by classical signal processing.
QSP achieves the manipulation of amplitudes of quantum states alternating a \textit{signal rotation} (a 2-dimensional operator, usually a rotation about $x$) and \textit{signal processing rotations} (usually about $z$).
To be more precise, let $W$ be the signal rotation about $x$
$$
    W = \begin{pmatrix}
        \cos(\theta) & i\sin(\theta) \\
        i\sin(\theta) & \cos(\theta)
    \end{pmatrix},
$$
then (as described in \cite[Theorem 1]{mrtc_unification_21}) for any given polynomials $P, Q\in\mathbb C[x]$ such that:
\begin{enumerate}[(i)]
    \item\label{prop:P_odd} either $P$ is even and $Q$ odd or vice versa;
    \item\label{prop:P_Q_norm} $|P(x)|^2 + (1-x^2)|Q(x)|^2 = 1$ for every $x\in[-1,1]$;
\end{enumerate}
there exists a sequence of real numbers $\Phi=(\phi_i)_{i=0}^{d}\in\mathbb R^{d+1}$ of length $d+1$, such that
\begin{itemize}
    \item $d$ is even if $P$ is even, it is odd if $P$ is odd;
    \item $d\geq\deg P$, $d>\deg Q$;
    \item the operator
    $$
        e^{i\phi_0\sigma_z}We^{i\phi_1\sigma_z}We^{i\phi_2\sigma_z}\ldots We^{i\phi_d\sigma_z}
    $$
    equals
    $$
        \begin{pmatrix}
            P(\cos(\theta)) & iQ(\cos(\theta))\sin(\theta) \\
            iQ^*(\cos(\theta))\sin(\theta) & P^*(\cos(\theta))
        \end{pmatrix}.
    $$
\end{itemize}
The converse holds as well.

QSVT is a sort of generalization of QSP obtained through a process called \textit{qubitization} \cite{Gilyen2019,Low2019}.
In order to keep the statement of QSVT as simple as possible, in the following, we restrict ourselves to the particular case we need in our discussion, that is, using real, even polynomial.


For any operator $\qop{O}$ and any projectors $\qop\Pi$ and $\qop{\tilde \Pi}$ on subspaces of the Hilbert space which $\qop{O}$ is acting on, and for any real vector $\Phi = (\phi_i)_{i=1}^d\in\mathbb R^d$, we define the operator $\qop{Q}=\qop{Q}(\qop{O},\Phi)$ --- acting on the same registers of $\qop{O}$ and on an auxiliary qubit in a register $\reg{B}$ --- as
\begin{equation}\label{eq:Q_expansion}
    \qop{Q} = \qop[B]{H}\qop{Q}_d\qop{Q}_{d-1}\ldots\qop{Q}_{1}\qop[B]{H}
\end{equation}
where each $\qop{Q}_j$ is defined as
\begin{equation}\label{eq:Q_j}
    \qop{Q}_j = \left\{\begin{array}{ll}
        \qop{R}_j\qop{O}    & \text{when $j$ is odd} \\
        \qop{\tilde R}_j\qop{O}^\dagger  & \text{when $j$ is even}
    \end{array}\right.
\end{equation}
    
and in turn $\qop{R}_j$ (and analogously $\qop{\tilde R}_j$) is implemented by a $\qop{X}$ gate controlled by $\qop{\Pi}$ ($\qop{\tilde\Pi}$ respectively), followed by a rotation about $z$ of angle $-2\phi_j$, followed again by the same controlled gate $\qop{X}$ (see also Figure \ref{fig:qsvt_block} for an example of $\qop{Q}_j$). 

Note that in this section and the next, we use the notation defined in Appendix \ref{app:notation}, and that Appendix \ref{subsec:qsvt} contains more details about QSVT.

Let $P$ be an even real polynomial of degree $d$ such that $|P(x)|\leq 1$ for all $x\in [-1,1]$. Since $P$ is even, there exists a real polynomial $\tilde P$ of degree $d/2$ such that $P(x) = \tilde P(x^2)$.
QSVT allows us to find a $d$-dimensional real vector $\Phi=(\phi_j)_{j=1}^d$ such that for every $\qop{O}$, $\qop{\Pi}$ and $\qop{\tilde\Pi}$ as previously defined, we have
\begin{equation}\label{eq:qsvt}
    \bra[B]{0}\qop{\Pi}\qop{Q}\qop{\Pi}\ket[B]{0} = \tilde P(\qop{\Pi}\qop{O}^\dagger\qop{\tilde\Pi}\qop{\tilde\Pi}\qop{O}\qop{\Pi}).
\end{equation}
In particular, when $\qop{\tilde\Pi}\qop{O}\qop{\Pi}$ is an $n\times n$, real and diagonal matrix, the right-hand side of equation \eqref{eq:qsvt} is still an $n\times n$, real and diagonal matrix whose entries are the evaluations of $P$ in the diagonal values of $\qop{\tilde\Pi}\qop{O}\qop{\Pi}$.


\section{Methodology}
\label{sec:methodology}
We now present an efficient methodology based on QSVT for building the state preparation operator needed to estimate VaR, assuming a multivariate Gaussian conditional independence model. 
It is important to note that our methodology can be extended to estimate CVaR, based on the procedure in Appendix \ref{subsec:expected_loss}.

Let us start with the settings of the problem.\\
Assume we have $n$ counterparties, that we enumerate from $1$ to $n$. The order of counterparties induces a natural enumeration of all possible scenarios of counterparty defaults: the $j$-th scenario, for $j$ between $0$ and $2^{n-1}$, will be the one for which the $i$-th counterparty defaults if and only if the $i$-th bit in the binary notation of $j$, which we denote by $j_i$, is $1$.
So, for example, in the $0$-th scenario no counterparty defaults, in the $(2^n-1)$-th scenario all counterparties default, in the $7$-th scenario the first and the third counterparties default.
The counterparties defaults might not be independent variables.
We denote the probability of the $j$-th scenario by $p_j$.

The study in \cite{Dri2023} demonstrated that in a multivariate setting, it is possible to build a circuit whose final state models the possible scenario.
To clarify, let $\qop{U}$ be the quantum circuit which encodes the probability distribution of all possible scenarios: it operates on an auxiliary register $\reg{Z}$,  used to encode a set of normally distributed random variables (the risk factors) to model the systemic risk\cite{Dri2023}, and on a quantum register of $n$ qubits, the so-called counterparties register $\reg{C}$.
The application of $\qop{U}$ results in
$$
    \qop{U}\ket[Z]{0}\ket[C]{0} = \sum_{j=0}^{2^n-1}\sqrt{p_j}\ket[Z]{\psi_j}\ket[C]{j}
$$
where every $\ket{\psi_j}$ is a normalized state vector in the auxiliary register $\reg{Z}$. In particular, the probability of obtaining $\ket{j}$ measuring the counterparties register is exactly $p_j$.

For every $i\in\{1,\ldots,n\}$ let $l_i$ denote the loss given default of the $i$-th counterparty, whereas for every $j\in\{0,\ldots,2^n-1\}$, we denote by $L_j$ the total loss in the $j$-th scenario, that is
$$
L_j  =\sum_{i=1}^{n} j_il_i.
$$
The maximal loss --- namely $L_{2^n-1}$ --- is also denoted $L_M$.

To avoid the use of the operators $\mathcal{S}$ and $\mathcal{C}$ and the full scan of the scenarios, we perform a two step process. 
We first map a function of the total loss directly in the amplitude of a qubit (register $T$) and then exploit QSVT \cite{Stamatopoulos2024} to apply a polynomial transformation on it. 
The transformation acts as a filter for the configurations satisfying the condition on the target loss (the total loss is lower than the target one).

\begin{figure*}
    \centering
    \begin{quantikz}
        \lstick[4]{$\ket[C]{j}$}&             &  \gategroup[5,steps=5,style={dashed,rounded corners,fill=blue!20},background,label style={label position=below,anchor=north,yshift=-0.2cm}]{$\qop{A}$}                             &   \ctrl{4} &      &   &  & \rstick[4]{$\ket[C]{j}$} \\
                                &             &                         &                         & \ctrl{3}                &         &  & \\
        \setwiretype{n}         &\vdots       &                         &                         &                         & \ddots  &  & \\
                                &             &                         &                         &                         &         & \ctrl{1}             & \\
        \lstick{$\ket[T]{0}$}   &             &                         \gate{\rotg{y}{2\theta_0}}  & \gate{\rotg{y}{2\theta(l_1)}} & \gate{\rotg{y}{2\theta(l_2)}} &\ \cdots\ & \gate{\rotg{y}{2\theta(l_n)}} & \rstick{$\cos(\alpha_j)\ket[T]{0}+\sin(\alpha_j)\ket[T]{1}$}
    \end{quantikz}
    \caption[The circuit of amplitude loading.]{The circuit of amplitude loading is the blue colored one. Each qubit in the counterparties register controls a rotation about $y$ whose angle encodes the loss of that same counterparty. Here $\alpha_j$ equals $\theta_0 + \theta(L_j)$.}
    \label{fig:amplitude_loading}
\end{figure*}
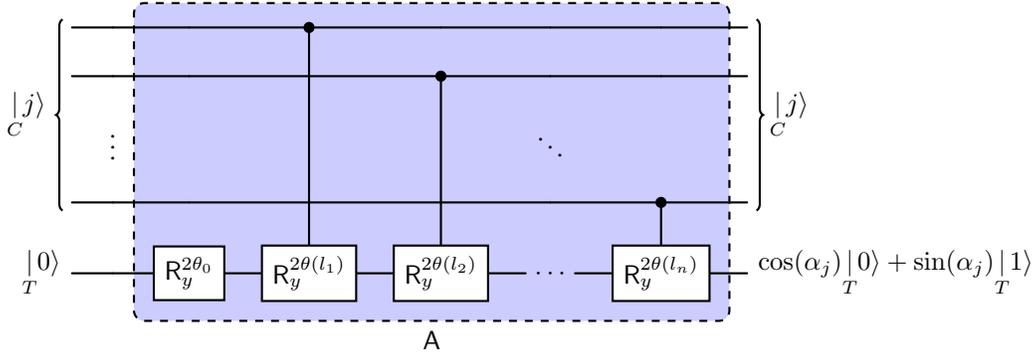
Let $\theta$ be an additive real function, that is
$$
\theta(\alpha + \beta) = \theta(\alpha) + \theta(\beta)
$$
for every $\alpha, \beta\in\mathbb R$.
We construct a circuit $\qop{A}$ acting on registers $\reg{C}$ and $\reg{T}$ by applying --- for every $i\in\{1,\ldots,n\}$ --- a rotation to $\reg{T}$ about $y$ of an angle $2\theta(l_i)$, controlled by the $i$-th qubit of $\reg{C}$ (see Figure \ref{fig:amplitude_loading}).
Since $\theta$ is additive, we have
$$
    \sum_{i=1}^n j_i\theta(l_i) = \theta\left(\sum_{i=1}^nj_il_i\right)= \theta(L_j)
$$
and therefore this implementation of $\qop{A}$ is such that
\begin{multline}\label{eq:application_of_amplitude_loader}
    \qop{A} \ket[C]{j}\ket[T]{0} = \\
    \cos\left(\theta_0 + \theta(L_j)\right)\ket[C]{j}\ket[T]{0} + \sin\left(\theta_0 + \theta(L_j)\right)\ket[C]{j}\ket[T]{1}
\end{multline}
and, moreover, linearly scales with respect to the number of counterparties (see Section \ref{subsec:depth}).


Now that we have a function that encodes the total loss in the amplitude, we can construct the transformation by checking the condition and thus filter the corresponding scenarios. 
For some real number $\mu\in(0,1)$ related to the target loss, consider the (even) threshold function $\thres{\mu}$ , that is
$$
\thres[x]{\mu} =\left\{\begin{array}{ll}
    1 & \text{when }|x|\leq\mu \\
    0 & \text{when }|x|>\mu
\end{array}\right..
$$
Since this function is even and has only two discontinuities in $\mu$ and $-\mu$, for any $\epsilon$ and $\delta$ positive real numbers, we can find an even polynomial $P=P_{\epsilon, \delta}$ such that
$$
    |P(x)-\thres[x]{\mu}| < \epsilon 
$$
for every $x\in(-1,1)$ such that $\lvert|x|-\mu\rvert>\delta$ and
$$
    |P(x)-\thres[x]{\mu}| \leq 1/2
$$
for every $x$ such that $||x| - \mu|\leq \delta$.
Of course, the smaller $\delta$ and $\epsilon$ are, the larger will be the required polynomial's degree.

We want to apply QSVT using such a polynomial, but unfortunately we cannot.
In fact, the best polynomial approximations would exceed $1$ in absolute value in the interval $(-1,1)$, violating one of the properties necessary for QSVT to be applied.
However, we can easily bypass this limitation by approximating $k\thres{\mu}$ for some proper positive real number $k<1$.
We will show that this workaround does not affect the final solution.

So, let us apply QSVT to $\qop{A}$ using:
\begin{itemize}
    \item an even polynomial of degree $d$ approximating $k\thres{\mu}$ in $[-1,1]$;
    \item $\qop{\Pi} = \ketbra[T]{0}{0}$ and $\qop{\tilde\Pi} = \ketbra[T]{1}{1}$ as projectors.
\end{itemize}
Note that
$$
\qop{\tilde\Pi}\qop{A}\qop{\Pi} = \sum_{j=0}^{2^n-1}\sin\left(\theta_0 + \theta(L_j)\right)\ketbra[C]{j}{j}\ketbra[T]{1}{0}
$$
that is represented by a diagonal real matrix.
Thus, applying QSVT we obtain $\qop{Q}$ --- acting on registers $\reg{C}$, $\reg{T}$ and on an auxiliary register $\reg{B}$ --- such that
\begin{multline}\label{eq:qsvt_applied_to_ampl}
    \ketbra[B]{0}{0}\qop[T]{\Pi}\qop{Q}\qop[T]{\Pi}\ketbra[B]{0}{0} = \\
    = \sum_{j=0}^{2^n-1}P\left(\sin\left(\theta_0 + \theta(L_j)\right)\right)\ketbra[C]{j}{j}\ketbra[T]{0}{0}\ketbra[B]{0}{0} \\
    \approx \sum_{j=0}^{2^n-1}k\thres[\sin\left(\theta_0 + \theta(L_j)\right)]{\mu}\ketbra[C]{j}{j}\ketbra[T]{0}{0}\ketbra[B]{0}{0} \\
    = \sum_{j\text{ s.t. } |\sin(\alpha_j)|\leq\mu}k\ketbra[C]{j}{j}\ketbra[T]{0}{0}\ketbra[B]{0}{0}
\end{multline}
where $\alpha_j= \theta(L_j) + \theta_0$ for every $j\in\{0,\ldots,2^n-1\}$.
Concatenating $\qop{Q}$ and $\qop{U}$ we obtain
\begin{multline}
    \qop{Q}\qop{U}\ket[Z]{0}\ket[C]{0}\ket[T]{0}\ket[B]{0} \approx \\
    \approx \sum_{j\text{ s.t. } |\sin(\alpha_j)|\leq\mu}k\sqrt{p_j}\ket[Z]{\psi_j}\ket[C]{j}\ket[T]{0}\ket[B]{0} + \ldots
\end{multline}
where the implicit remaining part is orthogonal to the subspace characterized by state $\ket{0}$ in registers $\reg{B}$ and $\reg{T}$.
Thus, the probability of measuring a state in such subspace is
\begin{equation}
\label{eq:cdf}
    \sum_{j\text{ s.t. } |\sin(\alpha_j)|\leq\mu}k^2p_j.
\end{equation}

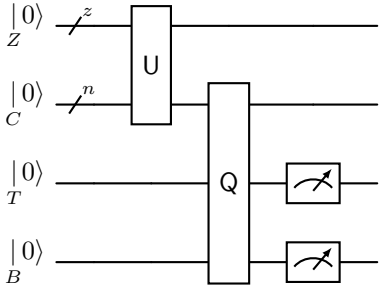
\begin{figure}
    \centering
    \begin{quantikz}
        \lstick{$\ket[Z]{0}$}&\qwbundle{z}  & \gate[2]{\qop{U}} &                   &           & \\
        \lstick{$\ket[C]{0}$}&\qwbundle{n}  &                   & \gate[3]{\qop{Q}} &           & \\
        \lstick{$\ket[T]{0}$}&              &                   &                   & \meter{}  & \\
        \lstick{$\ket[B]{0}$}&              &                   &                   & \meter{}  &  
    \end{quantikz}
    \caption{The circuit to encode the probability of having total loss at most some threshold in the probability of obtaining $0$ and $1$ measuring registers $T$ and $B$ respectively.}
    \label{fig:pre_amplitude_estimation}
\end{figure}

When $\theta$ and $\theta_0$ are properly chosen and in particular if
\begin{align}
    \sin\left(\theta\left([0,L_T)\right) + \theta_0\right)\subseteq [0,\mu) \label{eq:prop_less}\\
    \sin\left(\theta\left((L_T,L_M]\right) + \theta_0\right) \subseteq (\mu,1] \label{eq:prop_more}
\end{align}
for some $L_T\in(0,L_M)$, formula \eqref{eq:cdf} equals the cumulative distribution function evaluated in $L_T$, scaled by a factor $k^2$.

It remains to find a suitable function $\theta$.
A possible choice for $\theta$ could be
$$
\theta(x) = \frac{x\pi}{2L_M},
$$
accordingly choosing $\mu$ as $\cos(\theta(L_T))$.
However, in order to compute VaR using bisection search, this choice would require one to compute a different polynomial and QSVT phases at every step, since the threshold value would change.
 This procedure is a classical task that can be computationally intensive and time consuming, thus increasing the total preprocessing time required before running the quantum algorithm.
In fact, we opt for a different $\theta$ function that allows us to reuse a single polynomial even for different inputs of the problem.
A key point is that $\theta$ and $\theta_0$ must satisfy properties \eqref{eq:prop_less} and \eqref{eq:prop_more}, that translate in
\begin{align}
    \theta\left([0,L_T)\right) + \theta_0 & \subseteq [-\arcsin(\mu),\arcsin(\mu)),\\
    \theta\left((L_T,L_M]\right) + \theta_0 & \subseteq (\arcsin(\mu),\pi-\arcsin(\mu)].
\end{align}
However, since we are using an approximation of $\thres{\mu}$, a safety distance from $-\arcsin(\mu)$ and $\pi-\arcsin(\mu)$ should be kept, thus one may chose a minimal angle 
$$\beta_m\in(-\arcsin(\mu),\arcsin(\mu))$$
and a maximal angle 
$$\beta_M\in(\arcsin(\mu), \pi-\arcsin(\mu))$$
and make the image of $\theta(x) + \theta_0$ fit $[\beta_m, \beta_M]$.
This can be achieved by choosing $\theta$ such that
$$
    \theta(x) = x\frac{\arcsin(\mu) - \theta_0}{L_T}
$$
and $\theta_0$ satisfying
\begin{equation}
\left\{\begin{array}{lll}
    \beta_m &\leq\theta(0)&=\theta_0 \\
    \beta_M &\geq\theta(L_M)&=\frac{L_M}{L_T}\arcsin(\mu) - \frac{L_M}{L_T}\theta_0 + \theta_0
\end{array}\right.
\end{equation}
In order to improve performance of quantum estimation, the linear factor of $\theta$ should be as large as possible and consequently $\theta_0$ should be as smaller as possible, that is
\begin{equation}\label{eq:theta_0}
    \theta_0=\min\left\{\beta_m, \frac{L_M\arcsin(\mu)- L_T\beta_M}{L_M - L_T}\right\}
\end{equation}
Finally, aiming to keep the \textit{absolute value} of the multiplicative factor of $\theta$ as large as possible, one may even add a further degree of freedom, by possibly inverting $\beta_m$ and $\beta_M$ and using $\max$ instead of $\min$ in equation \eqref{eq:theta_0}. This will yield the tail distribution function, rather than the cumulative distribution function, but then it will be enough to complement the result in a post-processing phase in order to recover the cumulative distribution function.

However, our tests were run using the \textit{safest} conditions --- although probably not the best ---, adopting $\beta_m=0$, $\beta_M=\frac{\Pi}{2}$ and always adopting equation \eqref{eq:theta_0} as is.

\newcommand{\groupR}[1]{\gategroup[2,steps=3,style={dashed,rounded corners,fill=blue!20},background,label style={label position=above,anchor=north,yshift=-0.2cm}]{$\qop{R}_{#1}$}}
\newcommand{\groupQ}[1]{\gategroup[3,steps=4,style={dashed,rounded corners,fill=red!20},background,label style={label position=above,anchor=south,yshift=-0.2cm}]{$\qop{Q}_{#1}$}}

\subsection{Logical Depth of the Circuit}
\label{subsec:depth}

Let us focus on the logical depth of the circuit $\qop{QU}$ (see Figure \ref{fig:pre_amplitude_estimation}).
The complexity of the uncertainty model $\qop{U}$ has been already extensively studied in previous works \cite{Dri2022, Dri2023}.
Thus, we restrict our analysis to the depth of $\qop{Q}$, the remaining ingredient to build the circuit encoding the CDF.

\begin{figure*}
    \centering
    \begin{quantikz}
        \lstick{$\ket[C]{c}$}&\qwbundle{n}  & \gate[2]{\qop{A}}\groupQ{i} &                     &                        &                    & \gate[2]{\qop{A}^\dagger}\groupQ{i+1}&                      &                              &                &\\
        \lstick{$\ket[T]{t}$}&              &                             & \octrl{1}\groupR{i} &                        & \octrl{1}      &                                      & \ctrl{1}\groupR{i+1} &                              & \ctrl{1}       &\\
        \lstick{$\ket[B]{b}$}&              &                             & \gate{\qop{X}}      &\gate{\rotg{z}{2\phi_i}}& \gate{\qop{X}} &                                      & \gate{\qop{X}}       & \gate{\rotg{z}{2\phi_{i+1}}} & \gate{\qop{X}} &  
    \end{quantikz}
    \caption{Circuit for $\qop{Q_i}\qop{Q_{i+1}}$ (see equation \eqref{eq:Q_j}) in the particular case of our interest where QSVT is applied to $\qop{A}$ with $\qop{\Pi} = \ketbra[T]{0}{0}$ and $\qop{\tilde\Pi} = \ketbra[T]{1}{1}$. Here $i$ is intended to be odd}
    \label{fig:qsvt_block}
\end{figure*}
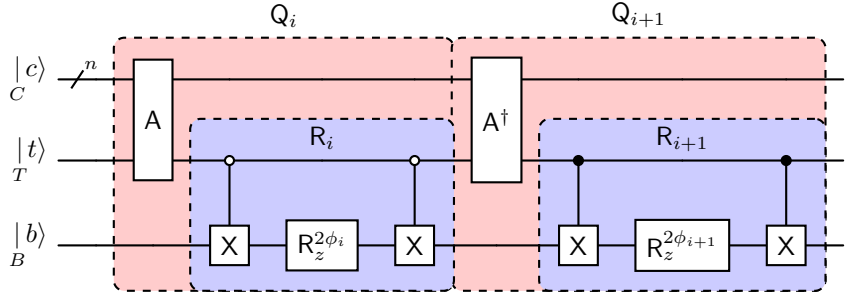

This is built by $d=\deg P$ instances of $\qop{Q_i}$'s. The depth of each $\qop{Q}_i$ does not depend on the parity of $i$ and is always given by the sum of the depths of each $\qop{R}_i$ --- that is a constant --- and the depth of $\qop{A}$, that linearly scales with the number $n$ of counterparties, since the amplitude loading operator is built concatenating $n$ controlled rotations and an uncontrolled one (see Figure \ref{fig:amplitude_loading}).

Putting everything together, we find that the depth of $\qop{Q}$ scales as
$O(dn)$.
Thus, for a fixed $d$, our method achieves linear scaling with respect to the number of counterparties $n$. The independence of $d$ from $n$ remains to be proved and will be analyzed in future works.

\newcommand{\aiqae}{\alpha_{\mathrm{IQAE}}}
\newcommand{\eiqae}{\epsilon_{\mathrm{IQAE}}}
\newcommand{\apcdf}{\mathrm{cdf}}

Once the circuit encoding the cumulative distribution function is built, the only difference between our solution and the original using a sum register \cite{Egger2021} is the presence of the factor $k^2$ which scales the value of the function in the amplitude (see equation \eqref{eq:cdf}).
When IQAE is applied to this circuit with a required confidence level $\aiqae$ and an error $\eiqae$, we obtain a value $v$ such that $k^2\apcdf$ is in $(v-\eiqae, v + \eiqae)$ with probability $\aiqae$.
Note that here $\apcdf$ denotes the value of the cumulative distribution function computed using the polynomial approximation of $\thres{\mu}$ to filter the possible scenarios.

In other words, we have found $\apcdf$ with error $\eiqae/k^2$.
Therefore, to compute $\apcdf$ with a desired error $\epsilon$, IQAE should be run with $\eiqae = \epsilon k^2$.
Since the asymptotic complexity of IQAE is $O\left(\eiqae^{-1}\log\left(\aiqae^{-1}\log_2\left(\eiqae^{-1}\right)\right)\right)$ (see \cite[equation (15)]{Grinko2021}), the presence of $k$ results in an increase of the amount of samples needed by a constant multiplicative factor.

\section{Results}
\label{sec:results}
To validate the proposed methodology, we present here the results of a simulation campaign for the end-to-end algorithm, from input to VaR computation. 
We also present an analysis of the outputs prior to the bisection search, namely the cumulative distribution function (CDF) calculated for a given value of aggregated loss.
This allows for assessing whether the quantum circuit implementing the algorithm indeed exhibits the expected behavior.

Table \ref{tab:input_data} contains the input values for the parameters of the model. The values correspond to the ones used in the experiments presented here, and are provided for reproducibility reasons.
Note that the values selected are not random but originate from a suitably anonymized and perturbed database containing real, not synthetic, data.

\begin{table}[ht]
    \centering
    \renewcommand{\arraystretch}{1.2}
    \setlength{\tabcolsep}{10pt}
    \begin{tabular}{ll}
        \toprule
        \textbf{Parameter} & \textbf{Value} \\
        \midrule
        \# of qubits per Gaussian & 2 \\
        Gaussian truncation values & $\pm2$ \\
        Number of assets  & 4 \\
        Intrinsic probability of default& $[0.256,\ 0.072,\ 0.135,\ 0.072 ]$ \\
        Sensitivity to risk factors ($\boldsymbol{\rho}$) & $[0.090,\  0.090,\  0.090,\  0.090]$ \\
        Loss given default (LGD) & $[ 18406.56,\ 54807.94,$ \\ 
                                                    &$\ 13719.59,\ 21127.25]$ \\\\
        
        Correlation with risk factors& $\begin{bmatrix} 
            0.158, & 0.058 \\
            0.256, & 0.157 \\
            0.158, & 0.058 \\
            0.158, & 0.058 
        \end{bmatrix}$ \\\\
        
        Number of shots ($n_{shots}$) & 2048 \\
        Confidence level ($\alpha_{var}$) & 0.05 \\
        \bottomrule
        \\
    \end{tabular}
    \caption{Input parameters for the experiments.}
    \label{tab:input_data}
\end{table}

The python code that implements our methodology and that was used to produce these results is available in \cite{repo}.

\begin{figure}
    \centering
    \includegraphics[width=1.1\linewidth]{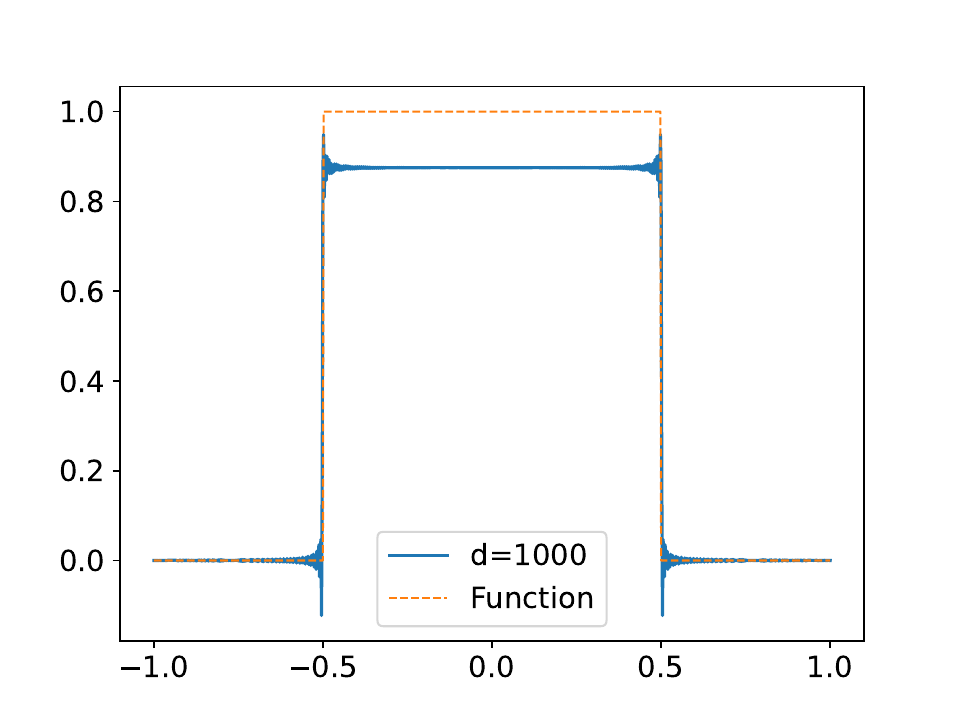}
    \caption{Polynomial approximation of the threshold function \(\thres{\mu}\) with degree \(d = 1000\), computed using the QSPPACK library. This approximation is employed in the QSVT framework.}
    \label{fig:poly_approx}
\end{figure}

Regarding QSVT, the computation of coefficients for the polynomial approximation of the threshold function and the estimation of phases for the operators were performed using the QSPPACK library\footnote{The open-source library can be found at \url{https://qsppack.gitbook.io/}.} that utilizes MATLAB.
The degree of the polynomial approximating the threshold function $\thres{\mu}$ was set to $d=1000$. The resulting polynomial is presented in Figure \ref{fig:poly_approx}.

\begin{figure}
    \centering
    \includegraphics[width=1\linewidth]{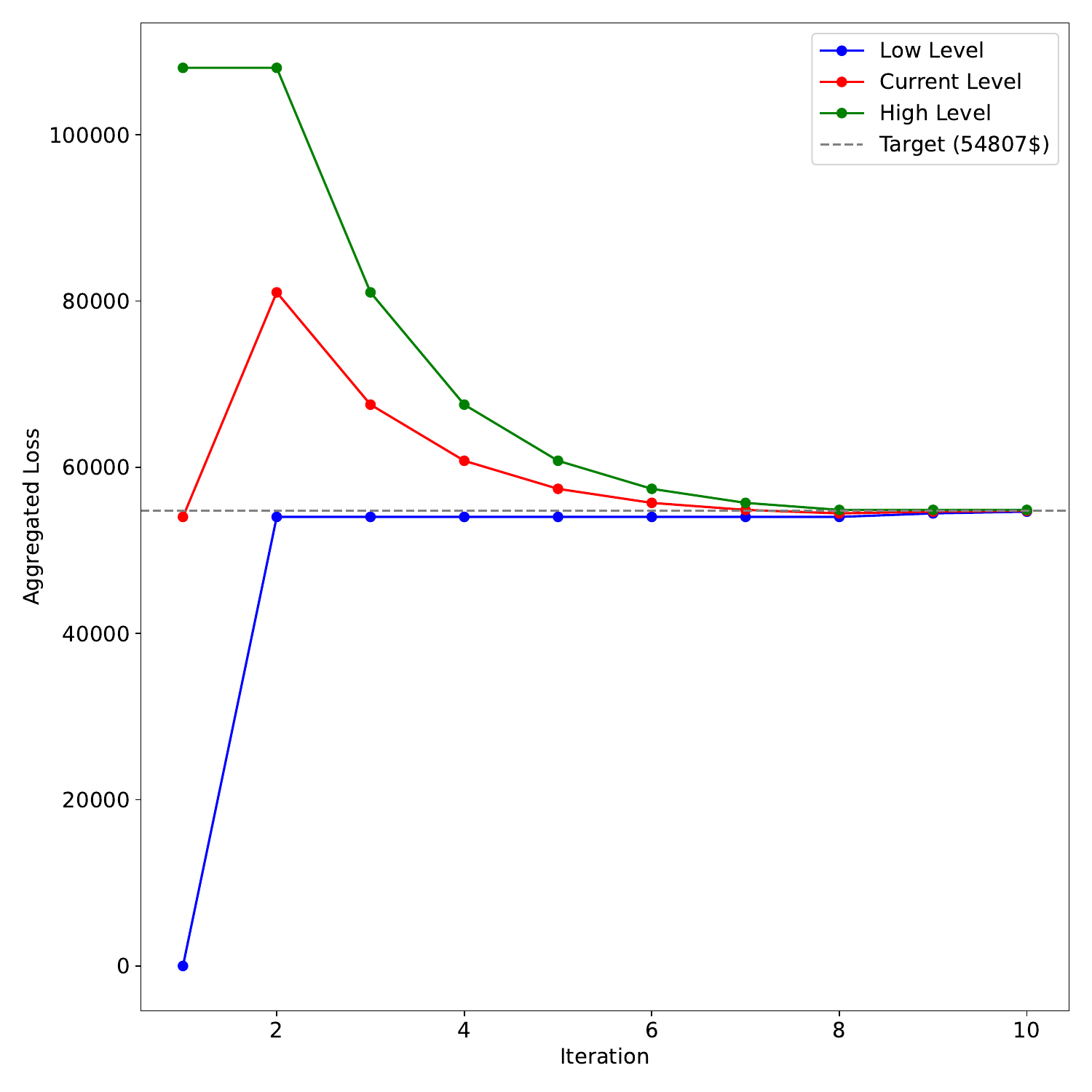}
    \caption{Illustration of the iterative bisection search process leveraging the quantum circuit for CDF estimation.}
    \label{fig:bisection_process}
\end{figure}

As mentioned above, to verify the effectiveness of the proposed circuit, the estimation of the CDF was tested for various values. The comparison is facilitated by computing the same estimate by simulating the original uncertainty operator circuit $\qop{U}$ to reconstruct the default probability for each possible scenario (that is also how the \textit{target} value in Figure \ref{fig:bisection_process} is obtained). 
From this, the CDF for each possible default scenario is then computed classically.
The reader can consult the result of this analysis in Table \ref{tab:cdf_values}.
Notice that values greater than one are due to the $k^2$ rescaling in post-processing (see Equation \ref{eq:cdf}).

\begin{table}[ht]
\centering
\caption{QSVT-based CDF benchmarking}
\label{tab:cdf_values}
\begin{tabular}{S[table-format=5.2] S[table-format=1.4] S[table-format=1.4]}
\toprule
{Target Loss} & \multicolumn{2}{c}{CDF} \\
\cmidrule(lr){2-3}
       & {QSVT} & {Benchmark} \\
\midrule
0         & 0.5827    & 0.5752 \\
13719.59 & 0.6676    & 0.6548 \\
18406.56 & 0.8457    & 0.8413 \\
21127.25 & 0.8905    & 0.8784 \\
32126.15 & 0.9128    & 0.9087 \\
34846.84 & 0.9228    & 0.9141 \\
39533.82 & 0.9350    & 0.9258 \\
53253.40 & 0.9369    & 0.9297 \\
54807.94 & 0.9735    & 0.9692 \\
68527.53 & 0.9800    & 0.9741 \\
73214.50 & 0.9957    & 0.9927 \\
75935.19 & 1.0019    & 0.9956 \\
86934.09 & 1.0018    & 0.9990 \\
89654.78 & 1.0065    & 1.0000 \\
94341.76 & 1.0008    & 1.0000 \\
108061.34& 1.0004    & 1.0000 \\
\bottomrule
\end{tabular}
\end{table}

The circuit used for estimating the CDF serves as the state preparation operator $\mathcal{A}$, which is subsequently employed by the QAE algorithm to achieve quadratic speed-up. 
Specifically, for the experiments, we adopted the iterative variant of amplitude estimation (IQAE), which offers the advantage of eliminating the need for a counting register, and consequently, additional qubits for phase estimation. Notably, IQAE does not require extra qubits and allows for direct specification of the precision parameter $\epsilon_{IQAE}$ and the confidence level parameter $\alpha_{IQAE}$, which determines the confidence interval for the final estimate.

Once the correct operation of the circuit in estimating the CDF is validated, it is then employed in the iterative bisection search process using the input data from Table \ref{tab:input_data}. 
As observed in Figure \ref{fig:bisection_process}, the proposed approach is indeed capable of converging towards the expected value (resulting from classically estimated calculations).

\section{Conclusion}
\label{sec:conclusion}
In this work, we presented a novel approach to quantum credit risk analysis that leverages QSVT to overcome the scaling challenges inherent in conventional QAE implementations. 
By replacing costly arithmetic operations within the quantum circuit with an amplitude loading strategy that scales linearly with the number of counterparties, our method circumvents the exponential gate complexity typically associated with similar approaches \cite{Dri2023}.

Our end-to-end implementation demonstrates that the proposed state preparation operator can accurately encode the uncertainty model, enabling effective estimation of the cumulative distribution function for aggregated losses. 

The simulation results validate the effectiveness of our methodology: the quantum circuit faithfully reproduces the expected CDF behavior and, when combined with a bisection search, converges towards the Value at Risk computed via classical Monte Carlo methods. 
These findings represent a step closer towards the possibility of efficiently realizing practical quantum risk analysis on quantum computers.

Future work could focus on further enhancements concerning the resources required for the scaling of the circuit related to state preparation.
It is also left to future work to study the details of the growth of the polynomial degree $d$ with respect to the required precision and the number of counterparties.
Furthermore, it would be beneficial to conduct a complete T-depth analysis to assess the effective benefit of our amplitude loading scheme in the fault-tolerant regime.

Finally, works concerning the extension of the proposed methodologies to other sophisticated uncertainty models of interest within the field of risk management could also be beneficial.

\section*{Acknowledgment}
The authors would like to thank Or Samimi Golan for the insightful discussions on QSVT.

\appendix


\subsection{Notation \& Conventions}
\label{app:notation}
In Section \ref{sec:qsvt} and \ref{sec:methodology}, when we need to write a linear operator acting on some Hilbert space, we write the name of this space or the quantum register associated with it below the operator itself. Thus, for example, as in equation \eqref{eq:Q_expansion}, we write $\qop[B]{H}$ to clarify that the Hadamard operator $\qop{H}$ is applied to the one-qubit register $\reg{B}$.
However, we omit this label whenever the Hilbert space is unambiguous.

In a somewhat similar manner, we may see kets as linear operators from $\mathbb C$ to a Hilbert space and, conversely, bras as linear operators from a Hilbert space to $\mathbb C$. So we write the name of the register under the vertical bar $|$. When a bra and a ket in the same space are concatenated, both the vertical bar and the space label collapse to one occurrence, that is, for example,
$$
\bra[B]{x}\ket[B]{y} = \braket[B]{x} y.
$$

From another point of view, we may consider operators as always acting on all quantum systems, but trivially, except for the one labeled under the operator name.
With this interpretation, we may naturally omit tensor products. We can also commute positions of operators acting on subspace whose intersection is trivial (for example when they act on different registers).

\subsection{A linear state preparation to compute the expected total loss and CVaR}\label{subsec:expected_loss}


Similar approaches to the one described in Section \ref{sec:methodology} allow to compute the expected total loss and Conditional Value at Risk (CVaR).
For example, CVaR can be calculated applying QSVT using a polynomial that approximates a function that inverts $\sin$ and applies a linear threshold function with some adjustment. That is, a function $\mathcal W_\mu$ such that
$$
    \mathcal W_\mu(x) = \left\{\begin{array}{cc}
        k\sqrt{\arcsin(\mu) - \arcsin(|x|)} & \text{for } |x|\leq \mu\\
        0 &  \text{for } |x|> \mu
    \end{array}\right.
$$
for a real positive $\mu\in(0,1)$, whose value depends on VaR, and a real scaling factor $k\in(0,1)$ that may be necessary to make the function feasible for QSVT.
However, following intuitions introduced in \cite{Woerner2019}, in this Section we present another solution that exploits the linear behavior of the first-order approximation of $\sin^2$ in $\pi/4$. 
In this way, we build a circuit to compute the expected total loss at first, and then, applying QSVT, we obtain CVaR.

Let $c$ be a small positive real number.
We chose in equation \eqref{eq:application_of_amplitude_loader} the function $\theta$ to be
$$
    \theta:x\mapsto \frac{cx}{L_M}
$$
and $\theta_0$ to be $\frac{\pi}{4}-\frac{c}{2}$.
In doing so, the probability of measuring $1$ in the register $\reg{T}$ is
$$
    P_1 = \sum_{j=0}^{2^n-1}p_j \sin^2\left(\frac{\pi}{4} - \frac{c}{2} + \frac{cL_j}{L_M} \right).
$$

By Taylor's expansion centered in $\pi/4$ we can approximate $\sin^2(x)$ as
\begin{multline*}
    \sin^2(x) = \\
    = \frac{1}{2} + (x - \pi/4) - \frac{4}{3!}(x-\pi/4)^3 + O(x-\pi/4)^5 = \\
    = \frac{1}{2} + (x - \pi/4) - \frac{2}{3}(x-\pi/4)^3 + O(x-\pi/4)^5 = \\
    = \frac{1}{2} + (x - \pi/4) + O(x-\pi/4)^3
\end{multline*}
and replacing $x$ by $\frac{\pi}{4} - \frac{c}{2} + c\frac{L_j}{\bar L_M}$, we obtain
\begin{multline*}
\sin^2\left(\frac{\pi}{4} - \frac{c}{2} + \frac{cL_j}{L_M}\right) = \\
=\frac{1}{2} + \frac{cL_j}{L_M} - \frac{c}{2} -\frac{2}{3}\left(- \frac{c}{2} + \frac{cL_j}{L_M}\right)^3 + O\left(- \frac{c}{2} +\frac{cL_j}{L_M}\right)^5\\
= \frac{1-c}{2} + \frac{cL_j}{L_M} -\frac{2}{3}\left(- \frac{c}{2} +\frac{cL_j}{L_M}\right)^3 + O\left(c^5\right).
\end{multline*}

Thus we can approximate $P_1$ as
\begin{multline}\label{eq:EL_P1}
P_1 =\\
=\sum_{j=0}^{2^n-1} p_j\left( \frac{1-c}{2} + \frac{cL_j}{L_M} -\frac{2}{3}\left(- \frac{c}{2} +\frac{cL_j}{L_M}\right)^3 + O\left(c^5\right) \right) \\
= \frac{1-c}{2} + \frac{c}{{L_M}}\sum_{j=0}^{2^n-1}p_jL_j -\frac{2}{3}\sum_{j=0}^{2^n-1}p_j\left(- \frac{c}{2} +\frac{cL_j}{L_M}\right)^3 + O\left(c^5\right) \\
= \frac{1-c}{2} + \frac{c}{{L_M}}\mathbb{E}[L] -\frac{2}{3}\sum_{j=0}^{2^n-1}p_j\left(- \frac{c}{2} +\frac{cL_j}{L_M}\right)^3 + O\left(c^5\right)
\end{multline}
whence
$$
\mathbb{E}[L] \approx \frac{{L_M}}{c}\left(P_1 - \frac{1}{2} + \frac{c}{2}\right).
$$
Thus applying IQAE we obtain the expected total loss.

To compute CVaR, instead, we may proceed similarly to what we did for VaR.
Given a real positive number $\mu\in(0,1)$, we consider a continuous variation of the linear threshold function, namely, a real function $\lthres{\mu}$ such that
$$
\lthres[x]{\mu}=\left\{\begin{array}{cc}
     \sqrt{-x^2+\mu^2} & \text{if }|x|\leq\mu \\
     0 & \text{if }|x|>\mu
\end{array}\right..
$$
Let $P$ be an even polynomial approximating $\lthres{\mu}$ (possibly rescaled by a factor).
Then equation \eqref{eq:qsvt_applied_to_ampl} in this configuration becomes
\begin{multline}\label{eq:qsvt_applied_to_ampl_cvar}
    \ketbra[B]{0}{0}\qop[T]{\Pi}\qop{Q}\qop[T]{\Pi}\ketbra[B]{0}{0} = \\
    = \sum_{j=0}^{2^n-1}P\left(\sin\left(\alpha_j\right)\right)\ketbra[C]{j}{j}\ketbra[T]{0}{0}\ketbra[B]{0}{0} \\
    \approx \sum_{j=0}^{2^n-1}\lthres[\sin\left(\alpha_j\right)]{\mu}\ketbra[C]{j}{j}\ketbra[T]{0}{0}\ketbra[B]{0}{0} \\
    = \sum_{j\text{ s.t. } |\sin(\alpha_j)|\leq\mu}\sqrt{\left(\mu^2 - \sin^2\left(\alpha_j\right)\right)}\ketbra[C]{j}{j}\ketbra[T]{0}{0}\ketbra[B]{0}{0}
\end{multline}
where
$$
\alpha_j = \theta_0 + \theta(L_j) = \frac{\pi}{4} - \frac{c}{2} + \frac{cL_j}{L_M}.
$$
So, concatenating $\qop{U}$ and $\qop{Q}$, the probability of measuring $0$ in both registers $T$ and $B$ is approximately
\begin{multline}
    P_1 = \sum_{j\text{ s.t. } |\sin(\alpha_j)|\leq\mu}p_j\left(\mu^2 - \sin^2\left(\alpha_j\right)\right) \\
    \approx \mu^2 - \sum_{j\text{ s.t. } |\sin(\alpha_j)|\leq\mu} p_j\left(\frac{1-c}{2} + \frac{cL_j}{L_M}\right) \\
    = \mu^2 - \frac{1-c}{2} - \sum_{j\text{ s.t. } |\sin(\alpha_j)|\leq\mu} p_j\frac{cL_j}{L_M}.
\end{multline}
Note that for $c<<1$, the sine of $\alpha_j$ is in the neighborhoods of $\frac{1}{2}$, thus the condition
$$
|\sin(\alpha_j)| \leq \mu
$$
may translate in
$$
\frac{\pi}{4} - \frac{c}{2} + \frac{cL_j}{L_M} \leq \arcsin(\mu)
$$
that is 
$$
L_j\leq \frac{L_M}{c}\left(\arcsin(\mu) - \frac{\pi}{4} + \frac{c}{2}\right)
$$
and one may choose $\mu$ such that the right-hand side of this last inequality equals the value at risk and therefore $P_1$ equals approximately $\mu^2 - \frac{1-c}{2} - c\frac{\mathrm{CVar}}{L_M}$.

\subsection{About QSVT}\label{subsec:qsvt}
In this paper we stated a result of QSVT that is not the one usually presented in literature.
Here we details steps from one of the main theorems of QSVT and our statement.
The main statement to which we refer is Corollary 11 in \cite{Gilyen2019}, but to present it, we must summarize some concepts.

Let $\mathcal H_O$ be a finite dimensional Hilbert space and let $\qop{O}$, $\qop{\Pi}, \qop{\tilde\Pi}\in\mathrm{End}(\mathcal H_O)$ be linear operators on $\mathcal H_O$ such that $\qop{O}$ is unitary and $\qop{\Pi},\qop{\tilde\Pi}$ are orthogonal projectors. Let $X$ and $\tilde X$ be the images of $\qop{\Pi}$ and $\qop{\tilde\Pi}$ respectively.
By singular value decomposition, there exist two orthonormal basis $\{\ket{\psi_i|i\in\{1,\ldots,\dim X}\}$ of $X$ and $\{\ket{\tilde\psi_i|i\in\{1,\ldots,\dim \tilde X}\}$ of $\tilde X$ such that
\begin{equation}\label{eq:svd}
    \qop{\tilde\Pi}\qop{O}\qop{\Pi} = \sum_{i=1}^{m}\zeta_i\ketbra{\tilde\psi_i}{\psi_i}
\end{equation}
where $m=\min\{\dim\tilde X, \dim X\}$ and $\zeta_1,\zeta_2\ldots,\zeta_m$ is an increasing sequence of non-negative real numbers. Such values are unique.
Whenever $P$ is an odd real polynomial, we write $P^{(SV)}(\qop{\tilde\Pi}\qop{O}\qop{\Pi})$ to mean $\sum_{i=1}^{m}P(\zeta_i)\ketbra{\tilde\psi_i}{\psi_i}$, whereas when $P$ is even $P^{(SV)}(\qop{\tilde\Pi}\qop{O}\qop{\Pi})$ stands for $\sum_{i=1}^{m}P(\zeta_i)\ketbra{\psi_i}{\psi_i}$.

\begin{theorem}[Corollary 11 in \cite{Gilyen2019}]\label{thm:qsvt}
    Let $\qop{O}$, $\qop{\Pi}$ and $\qop{\tilde\Pi}$ as previously defined.
    Suppose that $P\in\mathbb R[x]$ is a degree-$d$ polynomial satisfying that
    \begin{itemize}
        \item $P$ is an even function if $d$ is an even integer, it is odd if $d$ is odd and
        \item $|P(x)|\leq  1|$ for all $x\in[-1,1]$.
    \end{itemize}
    Then there exists $\Phi=(\phi_i)_{i=1}^d\in\mathbb R^d$ such that $P^{(SV)}(\qop{\tilde\Pi}\qop{O}\qop{\Pi})$ equals
    $$
       \left(\bra[B]{+}\qop{\tilde\Pi}\right)\left(\ketbra[B]{0}{0}\qop{U_\Phi} + \ketbra[B]{1}{1}\qop{U_{-\Phi}}\right)\left(\ket[B]{+}\qop{\Pi}\right)
    $$
    if $d$ is odd, and
    $$
        \left(\bra[B]{+}\qop{\Pi}\right)\left(\ketbra[B]{0}{0}\qop{U_\Phi} + \ketbra[B]{1}{1}\qop{U_{-\Phi}}\right)\left(\ket[B]{+}\qop{\Pi}\right)
    $$
    if $d$ is even,
    where $B$ is an auxiliary register and $\qop{U_\Phi}$ is the \emph{alternating phase modulation sequence}, that is defined as
    $$
         e^{i\phi_1(2\qop{\tilde\Pi} - I)}\qop{O}\prod_{j=1}^{(d-1)/2}\left(e^{i\phi_{2j}(2\qop{\Pi} - I)}\qop{O}^{\dagger}e^{i\phi_{2j+1}(2\qop{\tilde\Pi} - I)}\qop{O}\right)
    $$
    if $d$ is odd, and
    $$
        \prod_{j=1}^{d/2}\left(e^{i\phi_{2j-1}(2\qop{\Pi} - I)}\qop{O}^{\dagger}e^{i\phi_{2j}(2\qop{\tilde\Pi} - I)}\qop{O}\right)
    $$
    if $d$ is even.
\end{theorem}

We now focus on the case when $d$ is even.
Some straightforward algebraic computations show that
\begin{equation}\label{eq:E_j}
    \left(\ketbra[B]{0}{0}\qop{U_\Phi} + \ketbra[B]{1}{1}\qop{U_{-\Phi}}\right) 
    = \prod_{j=1}^{d/2}    \qop{\tilde E_{2j}}      \qop{O}^\dagger\qop{E_{2j+1}}\qop{O}
\end{equation}
where, for every $j\in\{1,\ldots,d\}$,
\begin{equation}\label{eq:Def_E_j}
\qop{E_j}=\left(\ketbra[B]{0}{0}e^{i\phi_{j}(2\qop{\Pi} - I)} + \ketbra[B]{1}{1}e^{-\phi_{j}(2\qop{\Pi} - I)}\right)
\end{equation}
and $\qop{\tilde E_j}$ is similarly defined replacing $\qop{\Pi}$ with $\qop{\tilde\Pi}$.
Let $\qop{\Pi^\perp}$ be the projector onto the orthogonal space $X^\perp$ of $X$. Then $I=\qop{\Pi} + \qop{\Pi^\perp}$ and therefore 
\begin{equation*}
    2\qop{\Pi} - I = \qop{\Pi} - \qop{\Pi^\perp}.
\end{equation*}
So, using matrix representation in a basis given by the union of a basis of $X$ and a basis of $X^\perp$, one can see that
\begin{equation}\label{eq:kickback}
    e^{i\phi_{j}(2\qop{\Pi} - I)} = e^{i\phi_j}\qop{\Pi} + e^{-i\phi_j}\qop{\Pi^\perp}
\end{equation}
that, in turn, this can be implemented by phase kickback.
Indeed, replacing equation \eqref{eq:kickback} into equation \eqref{eq:Def_E_j}, we obtain
\begin{align*}
    \qop{E_j} & = \left(e^{i\phi_j}\ketbra[B]{0}{0} + e^{-i\phi_j}\ketbra[B]{1}{1}\right)\qop{\Pi} \\
    & + \left(e^{-i\phi_j}\ketbra[B]{0}{0} + e^{i\phi_j}\ketbra[B]{1}{1}\right)\qop{\Pi^\perp} \\
    & = \rotg[B]{z}{-2\phi_j}\qop{\Pi} + \rotg[B]{z}{2\phi_j}\qop{\Pi^\perp}
\end{align*}
where $\rotg{z}{\alpha}$ denotes the rotation about $z$ of an angle $\alpha$. 

Now, denoting by $\qop{X}$ the NOT gate, observe that
$$
    \left(\qop{\Pi}\qop[B]{X} + \qop{\Pi^\perp}\right)\rotg[B]{z}{2\phi_j}\left(\qop{\Pi}\qop[B]{X} + \qop{\Pi^\perp}\right),
$$
which is our implementation of $\qop{R_j}$ as presented in Section \ref{sec:qsvt}, equals
\begin{equation}
     \qop{\Pi}\qop[B]{X}\rotg[B]{z}{2\phi_j}\qop[B]{X} + \qop{\Pi^\perp}\rotg[B]{z}{2\phi_j} \\
    = \qop{\Pi}\rotg[B]{z}{-2\phi_j} + \qop{\Pi^\perp}\rotg[B]{z}{2\phi_j} = \qop{E_j}
\end{equation}
since $\qop{X}\rotg{z}{2\phi_j}\qop{X} = \rotg{z}{-2\phi_j}$.
Similarly, it may be shown that $\qop{\tilde R}_j = \qop{\tilde E}_j$.

To fill the last gap needed to justify our statement of QSVT, let $P$ be the polynomial of Theorem \ref{thm:qsvt} in the even case and let $\tilde P(x)=\sum_{i=0}^{d/2}a_ix^i$ be the real polynomials such that $\tilde P(x^2) = P(x)$ for every $x\in\mathbb R$.
Then, using equation \eqref{eq:svd}
\begin{multline*}
    \qop{\Pi}\qop{O}^\dagger\qop{\tilde\Pi}\qop{\tilde\Pi}\qop{O}\qop{\Pi} =\\
    = \left(\sum_{i=1}^{m}\zeta_i\ketbra{\psi_i}{\tilde\psi_i}\right)\left(\sum_{i=1}^{m}\zeta_i\ketbra{\tilde\psi_i}{\psi_i}\right) \\
    = \sum_{i=1}^m \zeta_i^2\ketbra{\psi_i}{\psi_i}
\end{multline*}
and therefore
$$
\tilde P\left(\qop{\Pi}\qop{O}^\dagger\qop{\tilde\Pi}\qop{\tilde\Pi}\qop{O}\qop{\Pi}\right) = \sum_{i=1}^m \tilde P(\zeta_i^2)\ketbra{\psi_i}{\psi_i}= \sum_{i=1}^m P(\zeta_i)\ketbra{\psi_i}{\psi_i}.
$$
that equals $P^{(SV)}\left(\qop{\tilde\Pi}\qop{O}\qop{\Pi}\right)$ by definition.

\bibliographystyle{IEEEtran}
\bibliography{references}
\end{document}